\documentclass[12pt]{article}
\usepackage{epsf}
\usepackage{cite}
\def\hf{{1\over 2}}
\def\vecm{{\vec m}}
\def\roughly#1{\raise.3ex\hbox{$#1$\kern-.75em\lower1ex\hbox{$\sim$}}}
\def\kbdel{{K_{B\partial}}}

\def\calo{{\cal O}}
\def\op{{\omega_p}}
\def\ylm{{Y_{l\vecm}}}
\begin{document}
\bigskip
\hspace*{\fill}
\vbox{\baselineskip12pt \hbox{hep-th/9907204}}
\bigskip\bigskip\bigskip

\centerline{\Large \bf D3-brane shells to black branes on the Coulomb branch}
\bigskip\bigskip\bigskip

\centerline{\large Steven B. Giddings\footnote{\tt
giddings@physics.ucsb.edu}  and Simon F. Ross\footnote{\tt 
sross@cosmic.physics.ucsb.edu}}
\medskip
\centerline{Department of Physics,}
\centerline{University of California}
\centerline{Santa Barbara, CA 93106-9530}
\vskip 2cm

\begin{abstract}
We use the AdS/CFT duality to study the special point on the Coulomb
branch of ${\cal N}=4$ $SU(N)$ gauge theory which corresponds to a
spherically symmetric shell of $D3$-branes. This point is of interest
both because the spacetime region inside the shell is flat, and
because this configuration gives a very simple example of the
transition between D-branes in the perturbative string regime and
the non-perturbative regime of black holes. 
We discuss how this geometry is described in the dual gauge
theory, through its effect on the two-point functions and Wilson
loops. In the calculation of the two-point function, we stress the
importance of absorption by the branes. 
\end{abstract}

\newpage

The AdS/CFT correspondence \cite{mald:ads,gub:corr,witten:eucl} allows
us to study any asymptotically AdS geometry. Much work on this
correspondence has focused on learning about the field theory from
known properties of the supergravity, and hence has concentrated on
the study of perturbations on a pure AdS background. However, to study
the implications of the duality for gravitational physics, it is
interesting to consider a larger class of spacetimes.

A particularly interesting set of questions in the AdS/CFT
correspondence regards how the correspondence describes black holes.
We would especially like to understand how to treat black hole
formation, how to describe the observations of infalling observers and
compare them with observations made outside, and finally how (or
whether) the correspondence addresses issues of unitarity.

As an intermediate step on the path to such questions, one would like
to consider the simpler problem of the collapse of a shell of
three-branes to form a black brane.  There are static shell solutions
that are special cases of an interesting class of solutions
corresponding to multi-center $D3$-brane geometries where the
distribution of $D3$-branes has some special symmetry; that is, to
special points on the Coulomb branch of the gauge theory. Some
geometries of this type were obtained in \cite{kraus:coul}, as the
extreme limits of families of rotating $D3$-brane metrics. Solutions
with $D3$-branes distributed in disks of various dimensions were also
studied in \cite{free:coul,bran:coul}, where the two-point function
and Wilson loops in these backgrounds were discussed.

In the present paper, we will consider the geometry introduced in
\cite{kraus:coul}, where the $D3$-branes lie in a spherically
symmetric shell, so that the whole $SO(6)$ rotational symmetry of the
pure AdS$_5 \times S^5$ space is preserved. In this case, the metric
is
\begin{equation}\label{shmet}
ds^2 = H^{-1/2} (-dt^2+dx_{\parallel}^2) + H^{1/2} (dr^2 + r^2
d\Omega_5^2),
\end{equation}
where $H =1+ R^4/L^4$ for $r<L$, and $H=1+R^4/r^4$ for $r>L$. (Since
the spacetime is static, one could equally well consider a Euclidean
solution, but our main interest will be in the Lorentzian case.) In
the near horizon limit, which is $r,L\ll R$ in these coordinates, the
spacetime is the AdS$_5 \times S^5$ spacetime with curvature radius
$R$ outside the shell, and flat spacetime inside the shell. A
rescaling of the coordinates shows the near-horizon region includes
the whole AdS$_5 \times S^5$ spacetime. This near-horizon region will
be the main focus of this paper. Although the shell is at $r=L$, its
physical radius is $R$. Note also that the shell is flat in the
directions $x_\parallel$ which are along the individual branes; this
makes it possible to have a static configuration. We give the Penrose
diagram for this spacetime in Fig.~1. Unlike the geometries considered
in \cite{free:coul}, there is no region of large curvatures in this
spacetime. In fact, since the solution has no Weyl curvature, and the
local geometry is maximally supersymmetric, there should be no
$\alpha'$ corrections at all, by the argument of
\cite{kall:exact}. The solution as a whole will preserve only half the
supersymmetries; the other half are broken by the presence of the
$D3$-branes.

\begin{figure}
\leavevmode
\centering
\epsfysize=5cm 
\epsffile{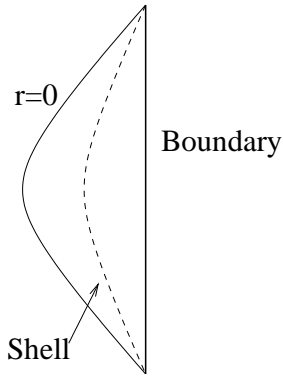}
\caption{Penrose diagram of the near-horizon spacetime with a
spherically symmetric shell. \label{fig1}}
\end{figure}

If one gives this solution a ``kick,'' it will collapse to form a
non-extremal black brane, as shown in Fig.~2. We have not yet been
able to analytically treat such solutions, so as a first step we will
instead consider the physics of the one-parameter family of static
solutions interpolating between large and small radius shells.  This
family provides a very simple example of a continuous path of
configurations that connects the perturbative and non-perturbative
regimes of string theory.  In one limit the shell becomes widely
separated $D3$-branes, easily treated by perturbative string theory, and
in the opposite limit the shell becomes a black brane with strong
gravitational effects.  It is here that the AdS/CFT correspondence and
related ideas have suggested profound modifications to our
semiclassical notions about black holes, and we are interested to get
more hints about how such modifications arise.

\begin{figure}
\leavevmode
\centering
\epsfysize=5cm 
\epsffile{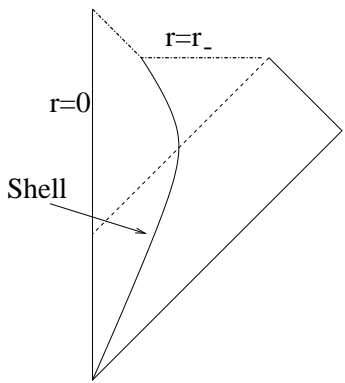}
\caption{A proposed Penrose diagram for the collapse of a three-brane
shell to form a non-extremal black three-brane. Note that the
singularity forms at $r=r_- >0$. \label{fig2}}
\end{figure}

Our interest in this solution has several other motivations. Firstly,
the metric is almost as simple as pure AdS, being static and having
the full rotational symmetry. This will make it possible to make
analytic calculations.  Secondly, it would seem to be interesting to
study how the flat space region in this geometry is described in the
gauge theory, in the hope that this might give more insight into the
extension of the correspondence to asymptotically flat boundary
conditions. (Unfortunately, we will not make much progress on this
issue.) Furthermore, analyzing a geometry of this kind forces us to
address the non-locality in the correspondence, since the geometry
outside $r=L$ is exactly AdS$_5 \times S^5$, and it will therefore not
signal its non-triviality in the one-point function $\langle {\cal O}
\rangle$ of any operator dual to a supergravity field.  Finally, these
configurations pose a new challenge to the AdS/CFT duality: as we'll
see, the supergravity predicts a spectrum of resonances that is not
yet understood from the Yang-Mills perspective.

The statement regarding vanishing of the one-point functions might
seem to be in conflict with the identification of this supergravity
solution with a point on the Coulomb branch of the gauge theory, since
the gauge theory would predict non-zero vacuum expectation values for
the Higgs fields $X^i$ (giving the positions of the branes in the
transverse space). However, the two points of view can be reconciled
so long as none of the operators which should have non-zero vevs from
the gauge theory point of view live in the short chiral primary
multiplets whose dimensions are protected. $\langle tr(X^i) \rangle$
will vanish by the spherical symmetry of the distribution. (At finite
$N$, it may have a value of $O(1/N)$, since the distribution doesn't
have exact spherical symmetry; this corresponds to a quantum effect
that will be invisible from our classical analysis.) An example of a
non-vanishing vev is $\langle tr(X^i X^i) \rangle \sim L^2$, but this
operator is not a chiral primary, so it should correspond to some
massive string excitation from the bulk point of view. It is
interesting that the gauge theory makes a definite prediction for the
value of this vev, but we were not able to use this to explore the
connection between this spacetime and the gauge theory.

As already remarked in \cite{dan:shell,vbsfr:part}, the presence of
the shell will affect the two-point function in the gauge theory, and
it should be possible to learn about how it is represented in the
gauge theory by studying this two-point function. The main point of
the present paper is to carry out this analysis of the two-point
function,\footnote{While this paper was in preparation,
\cite{chep:shell} appeared, which has a substantial overlap with our
calculation. One important distinction is that we will include the
effects of absorption by the shell.} and use it to probe the
properties of the shell configuration and the transition from
perturbative $D3$-branes to configurations with macroscopic horizons.
Another probe which will be sensitive to the presence of the shell is
a Wilson loop, and we will briefly discuss the effect of the shell on
the Wilson loops at the end of the paper.

According to the prescription of \cite{gub:corr,witten:eucl}, the
generating function for correlators in the gauge theory is identified
with the partition  function of supergravity, with
appropriately specified boundary conditions:
\begin{equation}\label{partfn}
\langle \exp i \int \phi_0 {\cal O} \rangle = Z[\phi_0].
\end{equation}
At large $R=(g^2N)^{1/4}$ (we set $l_s=1$), we can thus calculate the
two-point function of an operator dual to some test scalar field
propagating in the background (\ref{shmet}) in the semiclassical
approximation for supergravity, by calculating the action for the
scalar field subject to appropriate boundary conditions. For
convenience, we will work in momentum space, so what we want are
solutions to the ten-dimensional wave equation $\nabla^2 \phi = 0$ of
the form
\begin{equation}\label{soln}
\phi = e^{-i \omega t + i \vec{k} \cdot \vec{x}_\parallel} \Phi(r,\Omega).
\end{equation}
Since the spacetime is spherically symmetric in the $S^5$, there will
be no coupling between different partial waves.  In the flat region $r
< L$, the appropriate solution for $\Phi$ is
\begin{equation}\label{flat}
\Phi(r,\Omega) = {A \over r^2} 
J_{l+2} (q {R^2 \over L^2} r) Y_{l\vecm}(\Omega),  
\end{equation}
where $q^2 = \omega^2 - k^2$. In the AdS region, the solution is
\begin{equation}\label{ads}
\Phi(r,\Omega) =  - \left({q\over 2}\right)^{l+2} {\pi\over (l+1)! }
z^2  (N_{l+2} (q z) + C_l J_{l+2}(q z)) Y_{l\vecm}(\Omega),
\end{equation}
where we have introduced the convenient coordinate $z = R^2/r$ in the
AdS region. This solution behaves as $\phi \to z^{-l} e^{-i \omega t +
i \vec{k} \cdot \vec{x}_\parallel}Y_{l\vecm}(\Omega)$ as $z \to 0$,
which are the correct boundary conditions for it to give the
bulk-boundary propagator $\kbdel$.

If we neglect absorption by the shell, we should require that both
$\phi$ and its derivative are continuous at the shell. This fixes the
two constants of integration; in particular, it gives
\begin{equation}\label{C2}
C_l = - { J_{l+2}'(s) N_{l+2}(s) + J_{l+2}(s) N_{l+2}'(s) \over
2 J_{l+2}(s) J_{l+2}'(s)},  
\end{equation}
where $s = q R^2/L$ is a natural dimensionless measure of the
energy. As noted previously in \cite{free:coul,bran:coul,chep:shell},
the energy scale associated with the shell from the supergravity point
of view is $q \sim L/R^2 \sim L/\sqrt{gN}$, rather than the much
higher scale set by the Higgs vevs $q \sim L$. One suggestion for the
physics responsible for this scale is that it arises from states
consisting of W-boson-anti-W-boson pairs with large binding energy due
to the large coupling \cite{joe:priv}.

For general boundary conditions 
\begin{equation} \label{asymp}
\phi\rightarrow \int {d^4 k\over (2\pi)^4}
\sum_{l\vecm} \phi_{l\vecm}(k) z^{-l} e^{ikx} \ylm(\Omega),
\end{equation}
the action reduces to a surface term at infinity,
\begin{eqnarray}\label{action}
I &=& \hf \int_{z=0} d\Sigma^\mu \phi \nabla_\mu \phi \\
&=& {R^8\over 2} \int {d^4k \over (2\pi)^4} \sum_{lm} { d^4k' \over (2\pi)^4} 
\sum_{l'\vecm'} \phi_{l\vecm}(k) \phi_{l'\vecm'}(k') \lim_{z\rightarrow0}
z^{-l-3} \partial_z \kbdel(k,l,\vecm;k',l',\vecm';z). \nonumber
\end{eqnarray}
This gives a two-point function
\begin{equation}\label{2point}
\langle {\cal O}_{l\vecm}(\omega, \vec{k}) {\cal O}_{l'\vecm'}
(\omega',\vec{k}') \rangle = -i R^8 \lim_{z\rightarrow0} z^{-l-3} \partial_z
\kbdel(k,l,\vecm; k',l',\vecm';z).
\end{equation}
In evaluating the limit in this expression, there are singular terms which may
be dropped as they give contact terms on the boundary.  The final expression
is 
\begin{eqnarray}\label{2pt2}
\langle \calo_{lm}(\omega,k) \calo_{l'\vecm'}(\omega',k')\rangle &=&
  {iR^8 \over (l+1)! } \delta_{ll'} \delta_{\vecm\vecm'} (2\pi)^4
  \delta^4(k-k') \left( {q\over 2}\right)^{2l+4} {l+4\over (l+2)!} \\
&&\times  \left\{ 2\ln(q/2) - [\psi(1) +
    \psi(l+3)] + 2/(l+4) + \pi C_l \right\}. \nonumber
\end{eqnarray}

Notice that this two-point function has poles, at values given by the
zeroes of the denominator in $C_l$ (\ref{C2}).  The characteristic
spacing between these poles is given by the scale $L/R^2$ discussed
earlier; asymptotically the spectrum is given by $q \sim n \pi L/R^2$
for integer $n$. From the AdS perspective the presence of these poles
is understood by noting that the AdS plus shell physics is effectively
that of a cavity; the origin is regular, and there is a growing
potential at infinity.  This geometry will have normal modes, and the
poles in the propagator occur at the corresponding frequencies.  Note
that as the radius of the shell goes to zero, the geometry approaches
that of the charge-$N$ extremal three brane.  Allowed modes for this
form a continuum.  One can see this limit directly in the above; in
this limit the spacing between states vanishes.

Nonetheless, the $L\rightarrow0$ limit raises a puzzle, as it isn't
clear how this limit matches onto physics at $L=0$.  Indeed, to an
infalling observer, it might seem that these are very different
configurations: for $L\approx0$, the infalling observer would seem to
have a very painful impact on the shell resulting from the large
blueshift acquired on falling in, whereas the case $L=0$ results in a
smooth horizon. Of course, for any finite $L>0$, conformal symmetry is
broken in the boundary theory; from this perspective the limit isn't
necessarily expected to be smooth.\footnote{We thank Juan Maldacena
for a discussion on this point.} 

Another puzzle regards the emergence of the above sharp resonances.
It is not clear how this behavior smoothly goes over into the expected
$L=0$ behavior, and we furthermore do not understand their origin in
the boundary description. One proposal is that the spectrum of
resonances is a bound-state spectrum of the W-anti-W pairs mentioned
earlier. At tree level, the lightest W states have masses $\sim
g^{1/5}/R^{4/5}$, but tight binding effects could conceivably produce
a spectrum with the spacing $\sim L/R^2$ found earlier. 

We do not have a resolution to these puzzles, but part of the story
certainly involves an important physical effect that we have
neglected.  A string impinging on a single $D3$-brane may be absorbed,
converting into open string states on the brane.  Therefore, when we
send a wave into the shell geometry, it has a finite amplitude to be
absorbed by the shell.  In order to produce the resonant behavior
responsible for the above poles, the wave must pass through the shell
an infinite number of times, which it can't do without eventually
being absorbed.  In the presence of absorption the resonances are no
longer infinitely sharp.  We don't yet have an analytic calculation of
this effect, but we will give some estimates of its magnitude.

To begin, recall that if a wave has a frequency $\omega$ at infinity
in the full asymptotically flat solution (or in the pure AdS case, at
$r=R$), it will acquire the large blueshift noted above and have a
proper frequency $\omega_p = R\omega/L$ when it reaches the shell.
For small $L$ this frequency is quite high even for moderate
asymptotic frequencies.  At low frequencies we'll use known results to
estimate this absorption rate.  But for $\omega_p\roughly> 1$ in
string units, one expects the absorption cross section to climb
steeply as one encounters the increasing number of excited open string
states that may be created on the brane.  In this region it is
difficult to make precise estimates.

For frequencies $\omega_p<1$, we will estimate the absorption using
the results of \cite{kleb:3braneabs}.  Consider a plane wave of
frequency $\omega_p$ incident on a planar configuration of branes with
$n$ branes per unit area; for later correspondence with the shell
case, we will take $n\sim N/ R^5$.  The incident wave can create long
strings of length $\sim\omega_p$ connecting the branes.  This means
that we should think of a single cluster of branes with radius $\sim
\omega_p$ as a single absorber, and we will approximate the absorption
cross section for the cluster by the result for an equal number of
coincident branes.  This cluster has $N_{eff}\sim n \omega_p^5$
branes.  For $\op<1$ the effective radius $R_{eff} = (gN_{eff})^{1/4}$
satisfies $\op R_{eff}\ll1$, and so the cross-section, which grows
like $(\op R_{eff})^{8+4l}/\omega_p^5$ for the $l$th partial wave
\cite{magoo}, is dominated by the case $l=0$.  Thus the cross-section
for the cluster is
\begin{equation}\label{patchabs}
\sigma_{cl} \sim g^2 N_{eff}^2 \omega_{p}^3 \sim g^2 N^2 {
\omega_{p}^{10} \over R^{10}} \omega_{p}^3 \sim {
\omega_{p}^{13} \over R^2}.
\end{equation}
Assuming that the absorption cross-sections for the different clusters
add, this means that the sheet of three branes has an absorption
cross-section per unit area, or ``opacity'' given by
\begin{equation} \label{op}
{1\over \op^5} \sigma_{cl}\sim {\op^8\over R^2}\ .
\end{equation}
So long as the cluster size is small compared to the radius of the
sphere, this planar analysis should be a good approximation for
spherically symmetric shells. Since the cluster size $\sim \op$ is
less than the string scale, this is true in our case. 

We can describe this absorption with modified boundary conditions at
the plane. We must always require that the field itself is continuous
at the plane, but the absorption can be accounted for by introducing a
discontinuity in the normal derivative:
\begin{equation}\label{jump}
[ n^\mu \nabla_\mu \phi]  = -\alpha \phi
\end{equation}
with complex $\alpha$.  We use here a notation $[\Psi] \equiv \Psi
|_{r=l+\epsilon} - \Psi |_{r=l-\epsilon}$. From the point of view of
the equations of motion, we are adding a complex potential $\alpha
\delta(r-L)$ to the previous homogeneous equation. The jump in the
probability flux is
\begin{equation} \label{bc}
[n^\mu J_\mu]/\op = -2 \mbox{Im} (\alpha) |\phi|^2/\op\ .
\end{equation}
We will therefore take 
\begin{equation} \label{alpha}
\mbox{Im}(\alpha)\sim {\sigma_{cl} \over \op^4}.
\end{equation}
The precise value of the real part will not affect the qualitative
conclusions.  The solutions in the interior and exterior regions are
still given by (\ref{flat}) and (\ref{ads}), but the new matching
conditions will give a different value for $C_l$, namely
\begin{equation}\label{newC2}
C_l =-{ J_{l+2}(s) N'_{l+2}(s) + N_{l+2}(s) J'_{l+2}(s) - {\alpha L\over qR}
J_{l+2}(s) N_{l+2}(s) \over 2 J_{l+2}(s) J'_{l+2}(s) - { \alpha L \over qR}
J_{l+2}^2(s)  }\ .
\end{equation}
Furthermore, the action no longer reduces to just a surface term at
infinity, as the surface terms induced on $r=L$ by the bulk action
inside and outside the shell will fail to cancel as a result of the
discontinuity (\ref{jump}). The action will be
\begin{equation} \label{newact}
I = \hf \int_{z=0} d\Sigma^\mu \phi  \nabla_\mu
\phi -\hf  \int_{z=R^2/l} d\Sigma^\mu\phi [
\nabla_\mu \phi].
\end{equation}
This modifies the result (\ref{2pt2}) to give a two-point function
\begin{eqnarray}\label{newtwo}
\langle \calo_{lm}(\omega,k) \calo_{l'\vecm'}(\omega',k')\rangle &=&
  {iR^8 \over (l+1)! } \delta_{ll'} \delta_{\vecm\vecm'} (2\pi)^4
  \delta^4(k-k') \left( {q\over 2}\right)^{2l+4} \\ && \times \left(
  {l+4\over (l+2)!} \left\{ 2\ln(q/2) - [\psi(1) + \psi(l+3)]
  \nonumber+ 2/(l+4) + \pi C_l \right\}  \right. \\ && \left. + \alpha
  R {\pi^2 \over (l+1)!} \left[ N_{l+2}(s) + C_l J_{l+2}(s) \right]^2
  \right), \nonumber
\end{eqnarray}
with $C$ given in (\ref{newC2}).\footnote{This approach to the
calculation of the two-point function involves a surface term at the
shell. Nonetheless, the boundary two-point function should still be
the scaled limit of the bulk two-point function; the latter also
includes a contribution arising from absorption at the shell.}

Now $\alpha$ is complex, so its presence in the denominator of $C_l$
will move half of the poles in this expression off the real axis a
distance proportional to the absorption by the shell.  The remaining
poles should be moved to complex values once the finite thickness of
the shell is accounted for.  This decay of would-be normalisable modes
via absorption by the shell is described from the field theory point
of view as the decay of a composite object into pairs of gauge bosons.
The fractional width of the resonances are easily estimated to be
$\calo(\omega^7 R^4/L^7)$.  For low frequencies the resonances remain
extremely sharp and the last term on the right hand side of
eq. (\ref{newtwo}) is subdominant.  At increasing frequency or
equivalently decreasing $L$ they broaden, and we expect these
estimates to break down and the shell to become opaque once the proper
frequency at the shell reaches the string scale. One might hope to
match this absorption onto the extremal three brane case at $L=0$, but
this remains a puzzle. Since the seperation between the resonances is
larger than their width up to $\op \sim 1$, we can only hope to recover
the continuum characteristic of the $L=0$ solution above this scale;
note that because of the large blueshift, this corresponds to $\omega
\sim L/R$, and this energy scale is still lower than the Higgs scale
from the gauge theory point of view. 

We can also calculate the Wilson loops in the boundary gauge
theory. We expect that as the scale of the Wilson loop exceeds
$R^2/L$, the string worldsheet spanning the loop will pass from being
some minimal surface in the AdS part of the geometry to being
stretched straight from the Wilson loop to the $D3$-brane shell. In the
original pure AdS calculation, the energy of the minimal surface
between two static quarks seperated by $d$ was found to be
\cite{mald:wil}
\begin{equation}\label{adsE}
E = -{4\pi^2 R^2 \over \Gamma(1/4)^4 d}.
\end{equation}
The energy of a pair of strings stretching from the quarks to the
surface $r=L$ is
\begin{equation}\label{strE}
E = - {L \over \pi},
\end{equation}
because it is less than the energy of the strings stretching to $r=0$
which were used to fix the zero of energy in this calculation by this
amount. Hence we see that if $d > R^2/L$, the straight strings will
have less energy than the minimal surface. When $d \sim R^2/L$, the
minimal surface stretches down to $r_0 \sim L$. Thus, we see that
there is complete screening of the quark-anti quark potential for $d >
R^2/L$, as in the calculations for disks in
\cite{free:coul,bran:coul}.

A disappointing feature of our results is that we have been unable to
access the flat space region. For the two-point function, we can only
calculate in a regime at energies below the Higgs scale, and we expect
that at higher frequencies the shell will become effectively opaque,
and will prevent the probes from entering the flat region. Similarly,
in the Wilson loop calculation, when the loop is sufficiently large
for the string spanning it to reach the shell, it simply breaks and
terminates on the shell. Thus, it seems we cannot learn anything about
flat space holography from this solution. This failure may be regarded
as having been prefigured by the calculation in \cite{aki:flat}, where
it was observed that the energy/distance relation becomes flat in the
region of interest.

To summarize, the spherically symmetric shell (\ref{shmet}) provides
an extremely simple example of a supergravity solution corresponding
to the Coulomb branch of the gauge theory. However, because of the
high degree of symmetry of the solution, none of the one-point
functions of operators dual to supergravity fields can have non-zero
values; in particular, the non-zero Higgs vevs are related to
expectation values for operators whose dimensions are not protected,
corresponding to massive string fields on the gravity side. It would
be interesting if these expectation values could be estimated from the
gravity side; however, lacking a better control of the stringy
details, we have had to focus on studying this solution more
indirectly, through the supergravity two-point function and Wilson
loop expectation values. Our desire to treat the shell realistically,
as being composed of $D3$-branes, forces us to face a complication in
the two-point function calculations: the boundary conditions at the
shell must be chosen to express the absorption of incident wavepackets
onto the 3-branes. We can thus only calculate the two-point function
at very low frequencies.  At high frequencies one might hope to use a
WKB approximation, but lack of detailed understanding of absorption
presents an obstacle.  Taking absorption into account gives the poles
in the two-point function an imaginary part, indicating that they
correspond to unstable resonances decaying into gauge fields. The
calculation of the Wilson loop is more straightforward, and indicates
a sharp cutoff in the quark--anti-quark potential at separations $d
\sim R^2/L$, beyond which there is complete screening of the
potential, as in \cite{free:coul,bran:coul}.  Both effects can be
interpreted as part of the transition from perturbative D-brane
behavior to that of a large extremal black brane.  It will be
interesting to understand such crossover phenomena in more detail.

\bigskip
\bigskip
\centerline{\bf Acknowledgments}
\medskip

It is a pleasure to thank S. Gubser, A. Hashimoto, S. Kachru,
J. Maldacena, R. Myers, H. Ooguri, J. Polchinski, and W. Taylor for
discussions.  S.G. would like to particularly thank the theory group
at Berkeley and Lawrence Berkeley Labs, the Institute for Theoretical
Physics and Center for Mathematical Physics at the University of
Amsterdam, and the organizers of the Strings '99 conference, where
parts of this work was carried out.  S.G. is supported in part by DOE
contract DE-FG-03-91ER40618, and S.F.R. by NSF grant PHY95-07065.

\providecommand{\href}[2]{#2}
\begingroup\raggedright

\endgroup


\begin{thebibliography}{10}

\bibitem{mald:ads}
J.~Maldacena, ``The large N limit of superconformal field theories and
  supergravity,'' Adv. Theor. Math. Phys. {\bf 2} (1998) 231,
\href{http://xxx.lanl.gov/abs/hep-th/9711200}{{\tt hep-th/9711200}}.

\bibitem{gub:corr}
S.~S. Gubser, I.~R. Klebanov, and A.~M. Polyakov, ``Gauge theory correlators
  from noncritical string theory,'' Phys. Lett. {\bf B428} (1998) 105,
\href{http://xxx.lanl.gov/abs/hep-th/9802109}{{\tt hep-th/9802109}}.

\bibitem{witten:eucl}
E.~Witten, ``Anti-de {S}itter space and holography,'' Adv. Theor. Math. Phys.
  {\bf 2} (1998) 253,
\href{http://xxx.lanl.gov/abs/hep-th/9802150}{{\tt hep-th/9802150}}.

\bibitem{kraus:coul}
P.~Kraus, F.~Larsen, and S.~P. Trivedi, ``The Coulomb branch of gauge theory
  from rotating branes,'' JHEP {\bf 03} (1999) 003,
\href{http://xxx.lanl.gov/abs/hep-th/9811120}{{\tt hep-th/9811120}}.

\bibitem{free:coul}
D.~Z. Freedman, S.~S. Gubser, K.~Pilch, and N.~P. Warner, ``Continuous
  distributions of D3-branes and gauged supergravity,''
\href{http://xxx.lanl.gov/abs/hep-th/9906194}{{\tt hep-th/9906194}}.

\bibitem{bran:coul}
A.~Brandhuber and K.~Sfetsos, ``Wilson loops from multicenter and rotating
  branes, mass gaps and phase structure in gauge theories,''
\href{http://xxx.lanl.gov/abs/hep-th/9906201}{{\tt hep-th/9906201}}.

\bibitem{kall:exact}
R.~Kallosh and A.~Rajaraman,
``Vacua of M theory and string theory,''
Phys. Rev. D {\bf 58}, 125003 (1998)
hep-th/9805041.

\bibitem{dan:shell}
U.~H. Danielsson, E.~Keski-Vakkuri, and M.~Kruczenski, ``Spherically collapsing
  matter in AdS, holography, and shellons,''
\href{http://xxx.lanl.gov/abs/hep-th/9905227}{{\tt hep-th/9905227}}.

\bibitem{vbsfr:part}
V.~Balasubramanian and S.~F. Ross, ``Holographic particle detection,''
\href{http://xxx.lanl.gov/abs/hep-th/9906226}{{\tt hep-th/9906226}}.

\bibitem{chep:shell}
I.~Chepelev and R.~Roiban, ``A note on correlation functions in AdS(5) / SYM(4)
  correspondence on the Coulomb branch,''
\href{http://xxx.lanl.gov/abs/hep-th/9906224}{{\tt hep-th/9906224}}.

\bibitem{joe:priv}
J.~Polchinski, private communication.

\bibitem{free:corr}
D.Z.~Freedman, S.D.~Mathur, A.~Matusis and L.~Rastelli,
``Correlation functions in the CFT(d) / AdS(d+1) correspondence,''
Nucl. Phys. {\bf B546}, 96 (1999)
hep-th/9804058.

\bibitem{kleb:3braneabs}
I.~R. Klebanov, ``World volume approach to absorption by nondilatonic branes,''
  Nucl. Phys. {\bf B496} (1997) 231,
\href{http://xxx.lanl.gov/abs/hep-th/9702076}{{\tt hep-th/9702076}}.

\bibitem{magoo}
O.~Aharony, S.S.~Gubser, J.~Maldacena, H.~Ooguri and Y.~Oz,
``Large N field theories, string theory and gravity,''
hep-th/9905111.

\bibitem{mald:wil}
J.~Maldacena, ``Wilson loops in large N field theories,'' Phys. Rev. Lett. {\bf
  80} (1998) 4859,
\href{http://xxx.lanl.gov/abs/hep-th/9803002}{{\tt hep-th/9803002}}.

\bibitem{aki:flat}
A.~Hashimoto, ``Holographic description of D3-branes in flat space,''
\href{http://xxx.lanl.gov/abs/hep-th/9903227}{{\tt hep-th/9903227}}.

\end{thebibliography}
\end{document}